\documentstyle[prb,aps,twoside,rotate,epsf]{revtex}

\newcommand{\be}{\begin{equation}}
\newcommand{\ee}{\end{equation}}

\begin{document}
\draft

\title{
Fractal Analysis of Protein Potential Energy Landscapes
}
\author{D.A. Lidar,$^{a,b}$\footnote{Current address: Department of Chemistry,
The University of California, Berkeley, CA 94720, USA.} D. Thirumalai,$^{c}$
R. Elber$^{b}$ and R.B. Gerber$^{b,d}$} 
\address{
$^a$Department of Physics, The Hebrew University of Jerusalem, Jerusalem
91904, Israel\\
$^b$Department of Physical Chemistry and The Fritz Haber Center for Molecular
Dynamics, The Hebrew University of Jerusalem, Jerusalem 91904, Israel\\
$^c$Department of Chemistry and Biochemistry, Institute for Physical
Science and Technology, University of Maryland, College Park, Maryland
20742, USA\\
$^d$Department of Chemistry, University of California - Irvine, Irvine, CA
92697, USA\\
}

\maketitle

\begin{abstract}
The fractal properties of the total potential energy $V$ as a function of
time $t$ are studied for a number of systems, including realistic models of proteins
(PPT, BPTI and myoglobin). The
fractal dimension of $V(t)$, characterized by the exponent $\gamma$, is almost independent of temperature and
increases with time, more slowly the larger the protein. Perhaps the
most striking observation of this study is the apparent universality
of the fractal dimension, which depends only weakly on the type of
molecular system. We explain
this behavior by assuming that fractality is caused by a
self-generated dynamical
noise, a consequence of intermode coupling due to
anharmonicity. Global topological features of the potential energy
landscape are found to have little effect on the observed fractal
behavior.\\

\end{abstract}

\markboth{{Lidar, Thirumalai, Elber and Gerber}}{{Submitted Aug. 1998}}

\section{Introduction}
The energy landscape of proteins near the folding temperature is often
described as complex.\cite{Frauenfelder:94,Wolynes:95} That is, the landscape is expected to be rugged and
composed of many minima and maxima, separated by barriers of varying
heights, the result of frustration due to
conflicting interactions between different segments of the
protein. The complex nature of the underlying energy landscape has
been demonstrated using ergodic measures\cite{Straub:94} and by exact
enumeration of minima and transition states of a
tetrapeptide.\cite{Czerminski:90} 
 The goal of the present work is to
investigate this energy landscape under a variety of conditions using tools
from fractal analysis. The trajectories needed for analysis were
generated using ``MOIL'', a
molecular dynamics (MD) program
which allows realistic simulations of small proteins using
semi-empirical potentials.\cite{Elber:95} Using MOIL, the total potential energy $V(t)$ was calculated as a
function of time, for several proteins of choice (PPT -- Pancreatic
Polypeptide, BPTI -- Bovine Pancreatic Trypsine Inhibitor and
Myoglobin) at constant temperature. For comparison, we also simulated
poly-alanine and NaCl. The $V(t)$ curves so obtained under a variety
of choices of parameters, 
were then subjected to a fractal analysis. The main purpose was to 
correlate the resultant fractal dimensions with the topology of the
energy landscape, and to understand the connection to the relevant physical parameters. Both MOIL
and the method used to calculate the fractal dimension of the $V(t)$
curves are described in Sec.\ref{methods}.

There are several
reasons for using a fractal approach in the analysis of the $V(t)$
curves. Firstly, there is an intimate connection between ``roughness'' and
fractality: much is known about the generation of rough surfaces using fractal
algorithms and conversely, on the analysis of rough surfaces in fractal
terms.\cite{Family:91,Russ:94} In this sense the notion of the roughness of the energy landscape of
proteins naturally calls for a fractal analysis, which can {\em quantify} the
degree of roughness. Secondly, fractal dimensions
often show up as ``universal exponents'', and are related to the critical
exponents of phase transition theory.\cite{Stanley} Conversely, there is a certain
universality in protein dynamics, in the sense that most proteins exhibit a
folding transition and multi-exponential relaxation times. This suggests a
common feature in their energy landscapes. While the rapid transition to folded states in small proteins is
best rationalized by the existence of a dominant native basin of attraction (NBA) or a folding 
funnel,\cite{Wolynes:95,Thirumalai:96} the
relaxation phenomenon might well be related to the roughness of the energy
landscape, and hence to its fractal dimension. A ``universal'' fractal
dimension would be a strong characterization of what it is that is common to
different energy landscapes. Thirdly, establishing that the energy landscape
of a protein is indeed fractal would allow one to relate the subject of
protein dynamics to a whole body of work on dynamics on fractals, e.g.,
diffusion on a fractal substrate.\cite{Avnir:book}

The main problem in trying to characterize the roughness of a
high-dimensional energy landscape, such as that exhibited by a
protein, is the proper choice of a reaction coordinate. In this work
we have chosen {\em time} for this purpose. The
advantage is that time is a universal coordinate (unlike any of the
spatial coordinates), and one can
think of the protein potential energy dynamics in terms of this
coordinate as a random walk taking place on a rough substrate. Indeed,
the rate of sampling of conformation space in complex many-body
systems has been shown\cite{Straub:94} to be a diffusive process in
the energy space. The 
universality of the time-coordinate, however, can also be a drawback, in that it is not very
sensitive to the global topology of the energy landscape, but rather
to the local structure of minima. Indeed, as
will be presented in detail in Sec.\ref{results}, our results appear
to contain
information mostly on the role of {\em anharmonicity} in determining
the fractal features of the $V(t)$ curves. As such, the results
reported here reflect more on the general properties of dynamics in anharmonic
potentials than on protein dynamics per se. These and other issues
will be discussed and analyzed in detail in Sec.\ref{discussion}.
Conclusions and a summary are presented in Sec.\ref{conclusions}. Finally, to
establish a simple reference model, an
analytical solution for the fractal dimension of the $V(t)$ curve in
the Rouse model is presented in the appendix.

\section{Methods}
\label{methods}

\subsection{MOIL Simulations}
\label{MOIL}
The MOIL program has been described extensively in the literature.\cite{Elber:95} In a nutshell, it takes as
input the x-ray structure of a given protein from the Brookhaven Protein
Data Base (PDB), minimizes this structure, and runs molecular dynamics. MOIL makes it possible to
simulate the energetics, dynamics and thermodynamics of systems that vary in
size, up to several thousands of atoms. It accounts for a number of different
energy contributions: van der Waals and electrostatic (non-covalent); bonds,
angles, torsions and improper torsions (covalent):

\be
V = \sum V_{\rm vdW} + \sum V_{\rm electrostatic} + \sum V_{\rm bond}
+ \sum V_{\rm angle} + \sum V_{\rm torsion} + \sum V_{\rm improper} .
\label{eq:V}
\ee

\noindent The sums are over all interacting pairs of particles, and
the components are given as follows:

\be
V_{\rm vdW} = {{a_i a_j} \over r_{ij}^{12}} - {{b_i b_j} \over
r_{ij}^{6}} .
\label{eq:vdW}
\ee

\noindent $a_i$ and $b_i$ are constants that depend on the type of
atom $i$; $r_{ij}$ is the distance between atoms $i$ and $j$.

\be
V_{\rm electrostatic} = {{q_i q_j} \over {\epsilon r_{ij}}} ,
\label{eq:es}
\ee

\noindent where $q_i$ is the charge on atom $i$ and $\epsilon$ is the
dielectric constant.

\be
V_{\rm bond} = k_{ij} \left( r_{ij} - \langle r_{ij} \rangle \right)^2
,
\label{eq:bond}
\ee

\noindent where $k_{ij}$ is the bond force constant. $\langle x
\rangle$ denotes the equilibrium value of coordinate $x$.

\be
V_{\rm angle} = \kappa_{ijk} \left( \theta_{ijk} - \langle \theta_{ijk}
\rangle \right)^2 ,
\label{eq:angle}
\ee

\noindent where $\kappa_{ijk}$ is the angle force constant and
$\theta_{ijk}$ is the angle between two sequential bonds, i.e., bonds
connecting atoms $i,j$ and $j,k$.

\be
V_{\rm torsion} = \sum_{m=1} \alpha_{m,ijkl} \cos^m(\phi_{ijkl}) .
\label{eq:torsion}
\ee

\noindent Here $i,j,k,l$ are four sequentially connected atoms;
$\phi_{ijkl}$ is the angle between the planes defined by atoms $i,j,k$
and $j,k,l$; $\alpha_{m,ijkl}$ is a constant which depends on the types
of atoms $i,j,k,l$ and on the periodicity of the rotation $m$ (see
Ref.[$\!$\onlinecite{Elber:95}] for details). Finally,

\be
V_{\rm improper} = \mu_{ijkl} \left( \Phi_{ijkl} - \langle \Phi_{ijkl}
\rangle \right)^2 .
\label{eq:improper}
\ee

\noindent Here $j,k,l$ are all connected to atom $i$ and $\Phi_{ijkl}$
is the angle between the planes defined by atoms $i,j,k$ and $j,k,l$.

In the present work, we
considered only the time dependence of the {\em total} potential energy $V$ due to all these
terms. The time series for a number of systems were generated using numerical simulations employing either the 
canonical or microcanonical ensembles. As is customary, all simulations were preceded by a
short initial annealing (heating) period. In
the ``equilibration mode'' (used mainly for testing), the kinetic energies of all particles were
rescaled in order to preserve the constant temperature (canonical ensemble). In the more important
``no-equilibration mode'' (default unless mentioned otherwise), no such rescaling was attempted and the
system was allowed to reach thermal equilibrium at the specified
temperature (microcanonical ensemble).

The exponents characterizing the fractal nature of $V(t)$ reported below derive from averages over groups of
25 runs differering only in the initial random velocity
distribution. A number of tests were performed to ensure that our results are not artifacts of the particular 
integration schemes used. We checked that using more than 25 runs did not significantly
affect our results statistically. Between groups of 25 but for fixed molecule-type, the runs
differed, e.g., in temperature, addition of a water solvation shell, treatment of the terminus hydrogens,
bond-freezing, and deletions of residues. Most runs started at a
temperature of $T_i=$1K and heated in 1K per time-step increments to the final temperature $T_f$, at which
the rest of the simulation was run. The total duration of the simulations, once
$T_f$ was reached, was usually 10 or 25psec, corresponding to 20000 or
50000 time-steps respectively, of 0.5fsec each. Exceptions to these
conditions include some shorter and longer runs, with some variation
in time-step size. They will be discussed in detail below.

\subsection{Fractal Analysis}
The $V(t)$ curves were subject to a standard fractal analysis, with the
purpose of calculating their fractal dimension 

\begin{equation}
\gamma = 2-\alpha ,
\label{eq:ag}
\end{equation}
where $\alpha$ is the H\"{o}lder exponent, defined in the next section.

\subsubsection{General Theory}
Given any set of
points $E=\{(x,y)\}$, one can estimate their fractal dimension by a number of
(in principle) equivalent procedures. A popular one is the so-called
``Minkowski Sausage'',\cite{Mandelbrot,Falconer} where one draws a circle of radius $R$ centered at each
point in $E$, and calculates the area $A(R)$ of the union (``sausage'') of all
such circles. The fractal dimension is then found as

\begin{equation}
\gamma(E) = \lim_{R \rightarrow 0} \left[ 2-{\log[A(r)] \over \log(R)} \right]
\approx {\rm slope}\{\log(1/R), \log[A(R)/R^2] \}
\label{eq:gamma}
\end{equation}

\noindent where the second (approximate) equality follows by assuming that the
first equality holds for all $R$ and multiplying it throughout by
$\log(1/R)$. This assumption is equivalent to a linear relationship between
$\log(1/R)$ and $\log[A(R)/R^2]$, so that $\gamma$ is found as the slope of
the corresponding log-log plot. If such a linear relationship is not found, it
is taken in practice as an indication that $E$ is not a fractal
set. Conversely, if at least one order of magnitude of linearity is observed,
the convention is to consider $E$ a fractal set in that
range.\cite{Pfeifer-Avnir:book} It should be noted that rigorously speaking
the curve of a {\em function} is generally not a self-similar fractal, but rather a {\em
self-affine} one. The latter refers to those cases where the axes
have different units, so that the scaling might be different in the two
directions. Indeed, a (deterministic) self-affine curve satisfies the
following scaling relation:

\begin{equation}
h(t) = b^{-\alpha} h(b\,t) ,
\label{eq:SA}
\end{equation}

\noindent which holds only on average for random processes. The ``H\"{o}lder'', or self-affinity exponent 
$\alpha$, is $2-\gamma$ [Eq.(\ref{eq:ag})] if the curve is subjected to the Minkowski analysis of 
Eq.(\ref{eq:gamma}).\cite{Stanley:2,me:PRE96} An example is the Weierstrass-Mandelbrot
function:\cite{Mandelbrot} 
$h(t)=\sum_{n=-\infty}^\infty b^{-n \alpha} (1-\cos b^n t)$, which satisfies
Eq.(\ref{eq:SA}) and whose fractal dimension is 2-$\alpha$. An equivalent
characterization of statistically self-affine fractals is through the
correlation function, which satisfies:

\begin{equation}
C_h(t) = \langle [ h(t+t_0) - h(t_0) ]^2 \rangle \sim t^{2\alpha} .
\label{eq:C}
\end{equation}

\noindent This relation is expected to hold for durations shorter than a
typical correlation time $t_{c}$, found from:

\begin{equation}
t_{c} = {{\int dt\: t \Gamma(t)} \over {\int dt\: \Gamma(t)}} ,
\label{eq:Gamma}
\end{equation}

\noindent where $ \Gamma(t) = \langle h(t_0) h(t+t_0) \rangle - \langle
h(t_0)^2 \rangle $.

\subsubsection{Analysis of Time Series}
In the present case the (self-affine) set under consideration is:
$E=\{(t,V(t))\}_t$, and the fractal analysis we applied is a
numerically well-behaved version of Eq.(\ref{eq:gamma}): the
``$\epsilon$-variation method''. A full description of the method is
beyond the scope of this paper; for full details see
Ref.[$\!\!$\onlinecite{Dubuc}]. Briefly, Dubuc {\it et al.}\cite{Dubuc} introduced a
method particularly well suited for the evaluation of the
self-affinity exponent. They demonstrated that their method has the
most stable {\em local} exponent in comparison to a variety of other
methods, such as box-counting and power-spectrum. In their notation,
the $\epsilon$-variation of a function $f$ is $V(\epsilon,f) =
\int_0^1 v(x,\epsilon) dx$, where the $\epsilon$ oscillation
$v(x,\epsilon)$ is: $v(x,\epsilon) = \sup_{x' \in R_{\epsilon}(x)}
f(x') - \inf_{x' \in R_{\epsilon}(x)} f(x')$, and where
$R_{\epsilon}(x) = \{s \in [0,1]: |x-s|<\epsilon \}$. The
corresponding log-log plot for the calculation of the self-affinity
exponent is $\left[
\log(1/\epsilon),\log(V(\epsilon,f)/\epsilon^2)\right]$, with the
exponent given by 2--slope. We applied this analysis in order to
extract the self-affinity exponent (which we refer to simply as the
``fractal dimension'' and denote by $\gamma$) of our $V(t)$ curves.

\section{Results}
\label{results}

In this section we will present the findings from our numerical
simulations. A discussion will be postponed to the next section. 
We will start by presenting a sample of ``raw''
simulation results, then move on to a detailed presentation of each of
the three proteins, compare them amongst each other, perform a
comparison with several test cases, estimate the effect of time-scales
of different processes, and finally study the role of anharmonicity
{\it vs} harmonicity. All our MD simulations
started out from the folded state (as determined by X-ray), which we
subsequently minimized using the Powell algorithm.\cite{Elber:95} The protein was then
heated from 1K in increments of 1K at every time-step, with a random
initial velocity distribution, until the desired final temperature was
reached. Fractal analysis was performed on data from the constant
temperature period only. Dynamics on all three proteins was run for
25psec, with a time-step of 0.5fsec (with a few exceptions, to be
detailed below). Fractal analysis was performed on the full 25psec
trajectories and on the first 10psec of each such trajectory. This process
was repeated for a wide selection of temperatures in the range
1K-800K, with a focus on the room temperature regime.

\subsection{``Raw'' Results}

\begin{figure}
\hspace{6em}
\vspace{2em}
\epsfysize=6cm
\epsfxsize=11cm
\epsffile{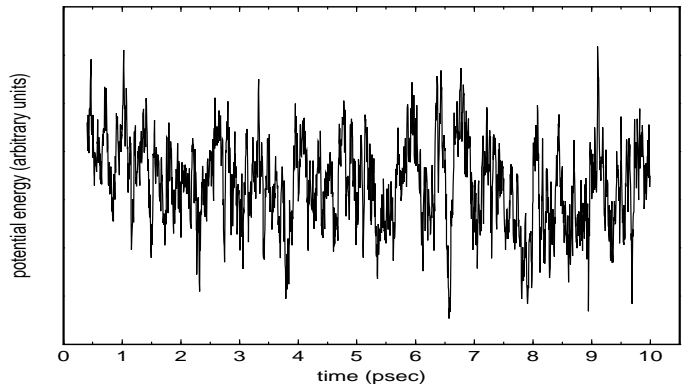}
\caption{Total potential energy as a function of time for a
single run of BPTI
at 310K. Annealing from 1K occurs during the first 0.4 psec, which are
not shown. The highly irregular fluctuations are reminiscent of
fractional Brownian noise.}
\label{fig:1}
\end{figure}

\begin{figure}
\hspace{6em}
\vspace{2em}
\epsfysize=6cm
\epsfxsize=11cm
\epsffile{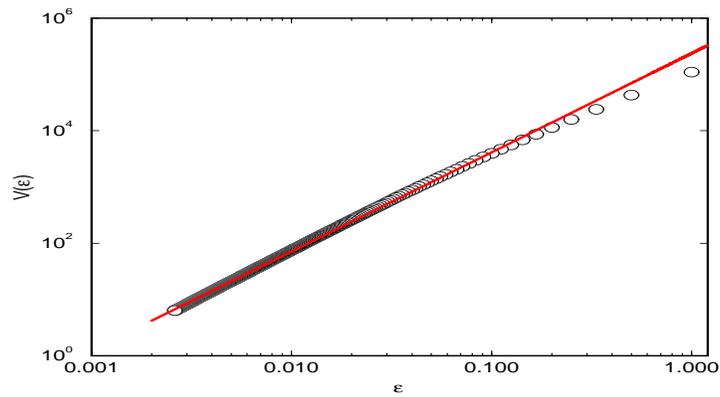}
\caption{$\epsilon$-variation\protect\cite{Dubuc} result for the curve
shown in Fig.\protect\ref{fig:1}. A straight line can be observed for about
1.5 decades. The bending at high $\epsilon$ values is due to finite
size effects.
}
\label{fig:2}
\end{figure}

\begin{figure}
\hspace{6em}
\vspace{2em}
\epsfysize=6cm
\epsfxsize=11cm
\epsffile{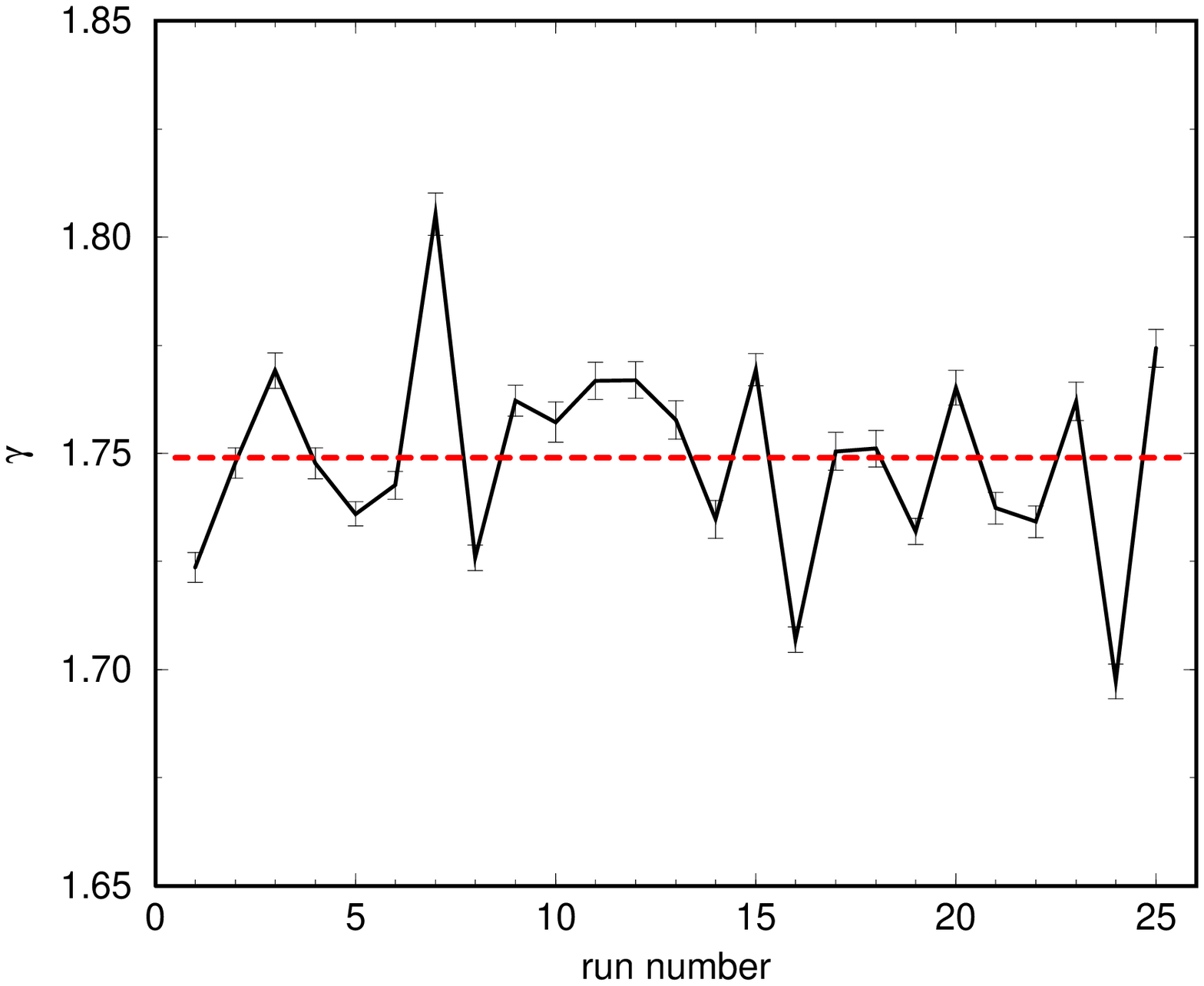}
\caption{All 25 fractal dimensions calculated for BPTI at
310K. Individual error bars are due to linear regression. The dashed
line is the average fractal dimension.
}
\label{fig:3}
\end{figure}

A graph of a typical $V(t)$ curve is displayed in Fig.\ref{fig:1}, for
BPTI at 310K. We only present the time series after the temperature is stabilized,
i.e., the first 0.4psec during which
annealing from 1K occurs are omitted. The noisy data are strikingly
similar to ``fractional Brownian motion'',\cite{Mandelbrot} and
clearly suggest the use of fractal-type analysis. This is done in
Fig.\ref{fig:2}, where the corresponding $\epsilon$-variation
analysis is shown. A straight line can be fitted with confidence to the
$\epsilon$-variation data over some 1.5 orders of magnitude (decades).
The linear regression coefficient in this case is: ${\cal R}^2 \!=\!
0.998$, where 1 
indicates a perfect straight line. The fractal
dimension (slope in Fig.\ref{fig:2}) is: $\gamma \!=\! 1.762\pm
0.004$, where the uncertainty is due to the linear regression. All log-log plots look strikingly
similar to the one in Fig.\ref{fig:2}, and have a similar range of
linearity. The ${\cal R}^2$ coefficient rarely went below 0.99. Fig.\ref{fig:3}
presents all 25 fractal dimensions $\gamma$ obtained from the runs for
BPTI under the abovementioned conditions. The dashed line is the
average fractal dimension, which in this case yields: $\gamma \!=\!
1.75 \pm 0.02$. The uncertainty in this number reflects both the
individual linear regression errors on each data point, as well the
standard deviation about the average. The procedure illustrated in
Figs.\ref{fig:1}-\ref{fig:3} was repeated throughout this work. Thus
every $\gamma$ data point in the figures to follow is the
result of an average over 25 fractal dimensions obtained from
independent runs differing only in the initial velocity distribution.

\subsection{Myoglobin}
\begin{figure}
\hspace{6em}
\vspace{2em}
\epsfysize=6cm
\epsfxsize=11cm
\epsffile{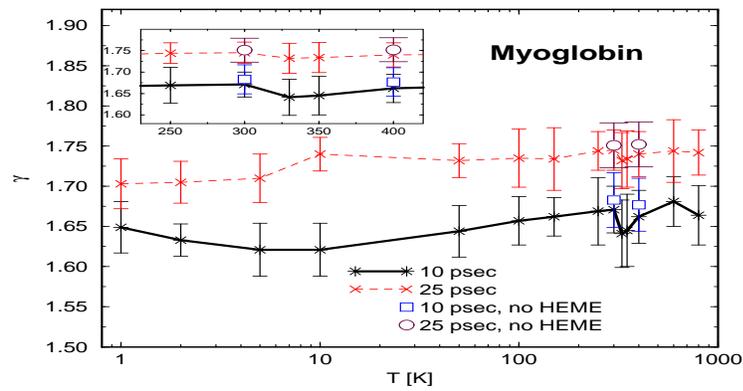}
\caption{Fractal dimension as a function of temperature for
myoglobin. Lines connecting data points are guides to the eye only.
}
\label{fig:4}
\end{figure}

\begin{figure}
\hspace{6em}
\vspace{2em}
\epsfysize=6cm
\epsfxsize=11cm
\epsffile{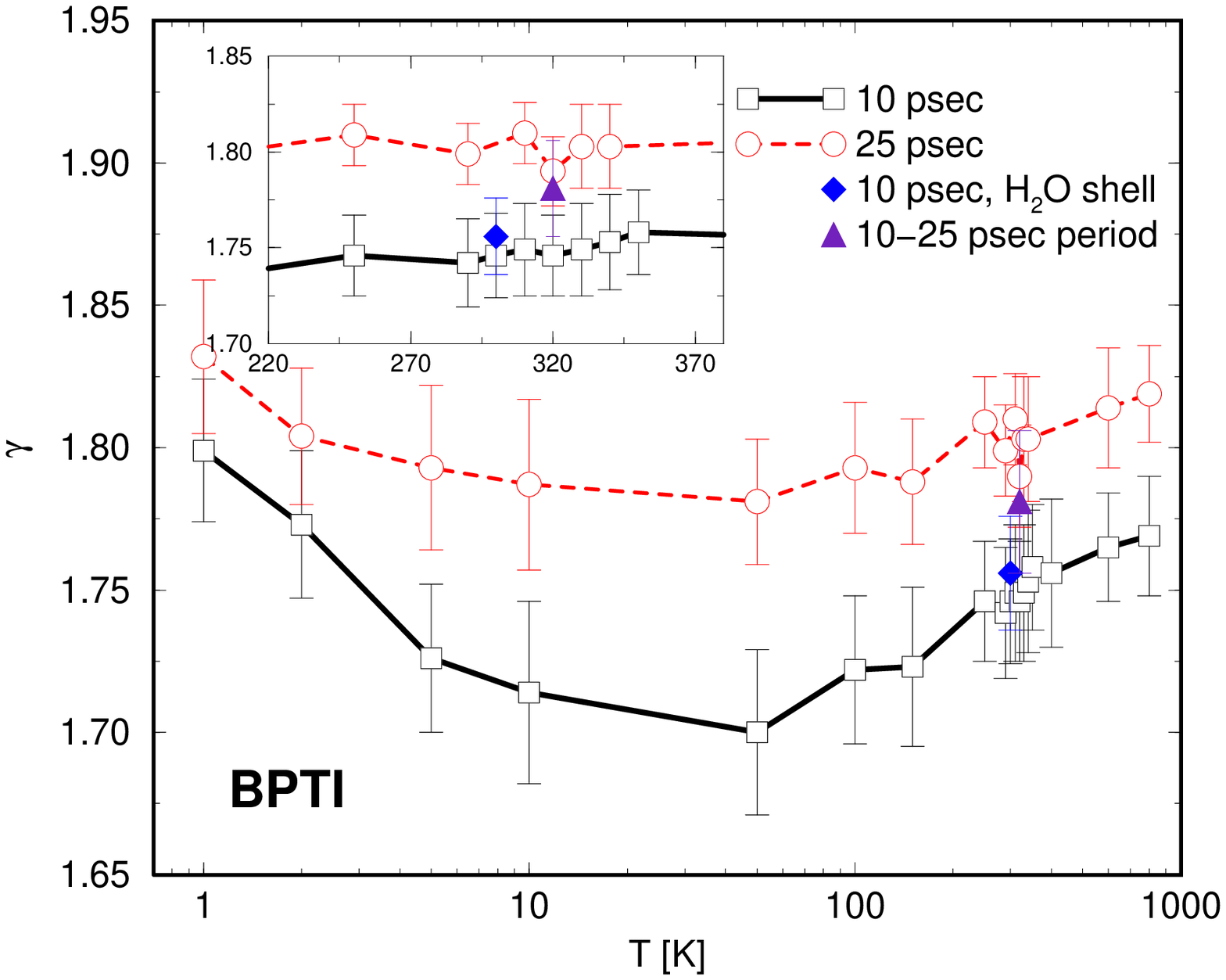}
\caption{Fractal dimension as a function of temperature for BPTI.
}
\label{fig:5}
\end{figure}

Fig.\ref{fig:4} shows the results of our simulations of
myoglobin. Every $\gamma$ point was obtained as discussed above, so
the error bars are due to both the averaging over the 25 data points
and the linear regression error on each of these points. The dashed (solid)
line connects the $\gamma$ values resulting from the 25psec (10psec)
runs. The inset shows the room temperature results. In the
10psec case $\gamma$ appears to have a shallow minimum around 10K,
which is however absent in the full 25psec simulations. The error bars
are in fact too large to confidently discuss a trend in the
curves. However, it is clear that {\em the fractal dimension {\bf
increases} with the simulation (or observation) time} $\tau$: on average $\gamma \!\approx\!
1.67$ for 10psec and rises to $\gamma \!\approx\! 1.72$ for
25psec. This behavior is seen in all our results
ahead (see Figs.~\ref{fig:5} and \ref{fig:6}).

As is clear from the inset, the fractal exponent $\gamma$ is essentially constant around
room temperature, apart from the rather sharp drop at 330K.

The robustness of the results was tested by running the simulations at
300K and 400K with the heme group deleted. The removal of the heme
group makes myoglobin considerably less compact, causing it to unfold
rather easily. Hence we expected a significant change in the
underlying dynamics. However, Fig.\ref{fig:4}
shows (squares and circles) that the influence of
heme deletion is very minor, and results in  a small
increase in $\gamma$.

\subsection{BPTI}
The results of our simulations of BPTI are shown in Fig.\ref{fig:5}. The
overall behavior is very similar to that of myoglobin. The differences
are (1) overall the fractal
dimensions are higher; (2) the
minimum is more pronounced (beyond statistical error), visible also
in the 25psec data, and has shifted to 50K.

The increase in $\gamma$ with increasing observation time for the longer simulation is unmistakable for
BPTI as well.

We also performed a 10psec simulation including a water
solvation shell (diamond). Clearly, this has no significant effect on
the results.

As a consistency test, we calculated $\gamma$ for the complementary
10-25psec of the simulation (triangle). The result is a
$\gamma$ value in between the 10psec and 25psec values. 

Finally, the dip observed for myoglobin can be seen here as well as in
the 25psec simulation, at T$=$320K.

\subsection{PPT}
Our PPT simulation results appear in Fig.\ref{fig:6}. The general
qualitative similarity to BPTI and myoglobin is noticeable, namely the
increase in overall fractal dimension with simulation time, and the
shallow minimum at around 50K. Several additional details are
noteworthy: (1) We included a 50psec long simulation at 1K (same
0.5fsec time step), and the increase in $\gamma$ persists (full
circle). (2) We also performed simulations
in the ``equilibration mode'' (see Sec.\ref{MOIL}). These are
indicated as squares and can be seen most clearly in the inset. The
results are virtually identical to the usual ``no-equilibration
mode'', demonstrating the agreement between the canonical and
microcanonical ensembles for large ($> 100$ atoms) systems. The only
exception is the point at 800K, where the 
``no-equilibration'' value of $\gamma$ drops rather sharply. (3) A water solvation shell
(diamonds, also in the equilibration mode) again does not produce
statistically significant results.

\subsection{Comparison Among Myoglobin, BPTI and PPT}
Figs.\ref{fig:7} and \ref{fig:8} compare the results for the three
proteins in the 10 and 25psec cases, respectively. These figures
merely recapitulate the results from Figs.\ref{fig:1}-\ref{fig:6} and
show no new data. The main points to notice are the striking
similarity between PPT and BPTI, compared to the overall lower fractal
dimension values for myoglobin. This holds in both the 10 
and 25psec cases.

\subsection{PPT Under a Variety of Conditions}
\begin{figure}
\hspace{6em}
\vspace{2em}
\epsfysize=6cm
\epsfxsize=11cm
\epsffile{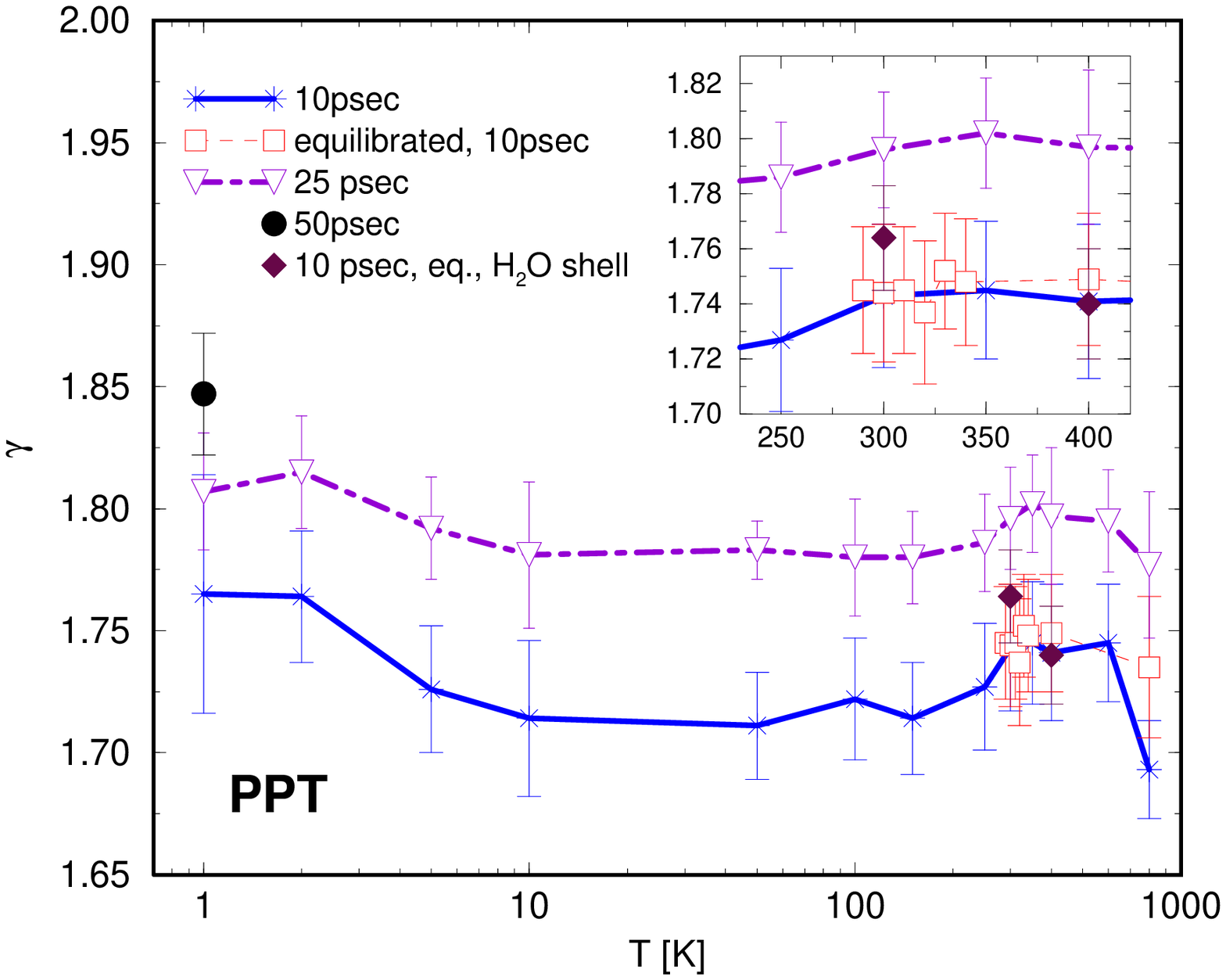}
\caption{Fractal dimension as a function of temperature for PPT.
}
\label{fig:6}
\end{figure}

\begin{figure}
\hspace{6em}
\vspace{2em}
\epsfysize=6cm
\epsfxsize=11cm
\epsffile{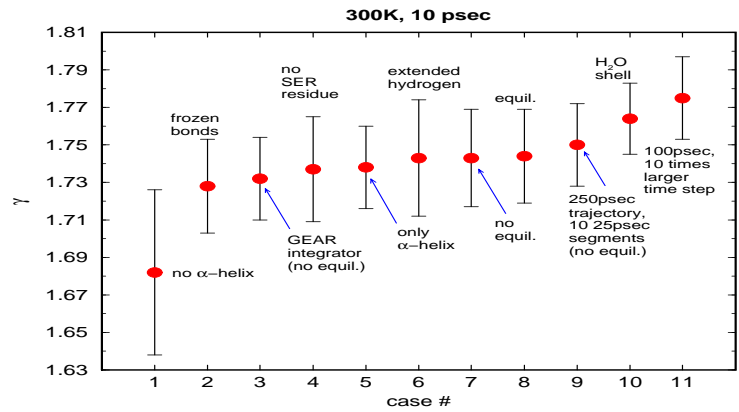}
\caption{A comparison of results for PPT at 300K, 10psec, under
a variety of conditions.}
\label{fig:9}
\end{figure}

We performed further tests of the robustness of our results by running
10psec simulations of PPT at 300K, under different conditions, as
shown in Fig.\ref{fig:9}. There are 11 data points in this figure, and the $\gamma$ values are
arranged in increasing order. Let us briefly describe the results,
noting that case 7 is the ``standard'' one (i.e., the result shown in
Fig.\ref{fig:6}).

The case ``no $\alpha$-helix'' refers to a
simulation in which the entire PPT $\alpha$-helix was deleted, i.e.,
only the GLY PRO SER GLN PRO THR TYR PRO GLY ASP ASP 
groups remain. Under these conditions
a significantly smaller $\gamma$ value is obtained. On the other hand
data point 5 is the opposite case, where only the $\alpha$-helix
remains. The corresponding $\gamma$ is much closer to the reference
value of data point 7. This shows {\em that the $\alpha$-helix is by and
large responsible for the observed fractal dimension}.

Data point 2
corresponds to a freezing of all bond vibrations. The remaining motions are
therefore proper and improper rotations and center of mass
translations. Since the deviation from the reference $\gamma$ value is
small, it appears that {\em the rotations are largely responsible for
the observed fractal dimension}.

Data point 3 is the result when a different integrator is used, which
appears to have only a minor effect. This test was performed to check
whether the results are a consequence of numerical instabilities.

Data point 4 is another deletion experiment, this time of just the SER
residue. This has a very small effect which, curiously enough, is slightly larger than leaving only the $\alpha$-helix, as
indicated in case 5 (although due to the large error bars in fact this
effect cannot be taken too seriously).

To calculate the $\gamma$ corresponding to data point 6, the terminal hydrogens were treated as
``extended'': unlike in all other cases, the hydrogens were not
treated as atoms but were ``absorbed'' into the carbons. This modifies
the carbon mass and radius to appropriate effective values. The effect,
however is negligible.

Data points 8 and 10 relate to using the equilibration mode and adding a
water solvation shell respectively, which (as shown also in
Fig.\ref{fig:6}) do not have a significant effect either.

Data point 9 reports the result from a single 250psec long trajectory: the
trajectory was divided up into 10 segments of 25psec each, and the
$\gamma$'s were averaged. Data point 11 is due to a 100psec long trajectory which, however, employed
a time step of 5fsec (10 times larger than the usual time step).  In accordance with the observation
of a fractal dimension which increases over time, the resulting
average $\gamma$ of data point 9, as well as that of data point 11, are larger
than the reference value.

\subsection{PPT {\it vs} Test Systems: Poly-alanine and NaCl}
\begin{figure}
\hspace{6em}
\vspace{2em}
\epsfysize=6cm
\epsfxsize=11cm
\epsffile{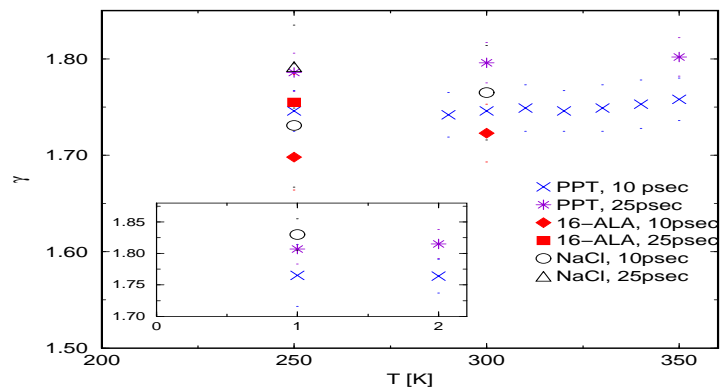}
\caption{A comparison between PPT, 16-alanine and NaCl (40
atoms). Error bars have been omitted for clarity and are replaced by
small dots.}
\label{fig:10}
\end{figure}

In order to estimate to what extent our results reflects properties
unique to realistic proteins, we present calculations for two test
systems: poly (16)-alanine and a 40-atom cluster of NaCl. Poly-alanine
folds into an $\alpha$-helix, so it has the crucial part of the
protein topology (recall points 1 and 5 in Fig.\ref{fig:9}), but the underlying energy landscape is not expected to be rugged. This is becuase all the residues are identical so there is no frustration due to conflicting interactions
between different segments of the polymer. (Nevertheless each residue
has internal conflicting interactions, giving rise to some
ruggedness). The ground state of the non-polymeric NaCl cluster is
easily accessible,\cite{Berry} i.e., starting from a disordered state, it finds it crystalline ground state well before
the end of our 10psec simulation. Thus it has neither the protein
topology, nor its rugged energy landscape. The results of our simulations
are shown in Fig.\ref{fig:10}. The 10 and 25psec runs of 16-alanine
and NaCl at 250K, as well as 10psec at 300K, and 10psec for NaCl alone
at 1K. For clarity of presentation the error
bars have been removed. However, it should be remarked that when the
error bars are taken into account the results for a given temperature
overlap. Indeed, it is striking that {\em the results for poly-alanine and
NaCl do not differ significantly from those for PPT}. In fact for
250K, NaCl and PPT agree almost exactly, and the agreement for 10psec
at both 250K and 300K is quite close. Only at 1K does there seem to be
a significant difference between NaCl and PPT. 16-alanine has lower
$\gamma$ values than PPT, but not very much so. Barring accidental
agreement, these results clearly
indicate that {\em neither the protein topology, nor the rugged energy
landscape are essential in order to obtain fractal $V(t)$ curves, with
fractal dimension close to that of realistic proteins}.

\subsection{The Role of Observation Time: Short Simulations of PPT}
\begin{figure}
\hspace{6em}
\vspace{2em}
\epsfysize=6cm
\epsfxsize=11cm
\epsffile{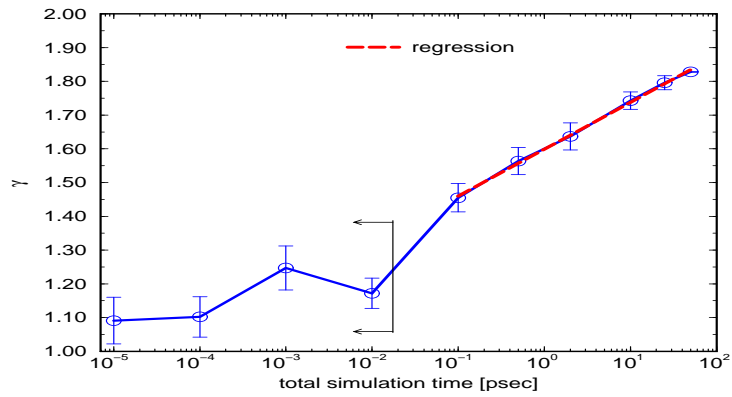}
\caption{20000 time step simulations of PPT at 300K with
varying total duration (x-axis). Results to the left of the arrows are
in the unphysical range.}
\label{fig:11}
\end{figure}

In order to systematically investigate the role of the total
duration of the simulation and the correspondingly participating
physical processes, we ran simulations of 20000 time steps of PPT
at 300K, with progressively smaller
time steps. This is different from the bulk of our simulations where a
constant time step of 0.5fsec was employed, so care should be
exercised in the comparison. The main point is that by employing a
smaller time step, previously inaccessible fast motions are now
observable. The results are shown in Fig.\ref{fig:11}. It is
remarkable that the fractal dimension obeys an almost perfect
logarithmic law over three orders of magnitudes: 

\begin{equation}
\gamma(\tau) = 1.60 + 0.060 \ln(\tau) , 
\label{eq:gamma-tau}
\end{equation}
for 0.1psec $ \le \tau \le $ 100psec. The
regression coefficient is ${\cal R}^2 \!=\! 0.999$. 
Fig.\ref{fig:11} thus demonstrates decisively
that $\gamma$ is indeed strongly dependent on the total observation time.\cite{comment} 

Below $\tau \!=\!$ 100fsec (to the left of the arrows in Fig.\ref{fig:11}) the behavior of
$\gamma$ appears rather erratic and we believe this is due to
numerical error. Indeed, the fastest motions in the protein are the OH
or CH stretches, with a period of about 10fsec. Thus $\tau \!=\!$
10fsec would most likely not probe even a single period and can
confidently be regarded as the lower physical limit of our
simulations.

We turn next to an analysis and discussion of our results.

\section{Discussion}
\label{discussion}

The most striking observation presented in the previous section is the
rise of the fractal dimension with time. It suggests (at least)
two likely hypotheses for its explanation: {\em As time evolves}

\begin{itemize}
\begin{enumerate}
\item{
there is increased access to local minima and ``fine-structure'' of the
potential energy landscape;}
\item{
low frequency vibrations and energy transfer between modes become
increasingly activated.}
\end{enumerate}
\end{itemize}

The first of these attributes the growth in $\gamma$ over time to
the increased roughness of the sampled potential energy landscape. In other
words, the protein ``visits'' more and more local minima (or basins of attraction) as its
samples its conformational space. This process is consistent
with an increase in fractal dimension since the fractal potential
energy landscape is not sampled all at once by the protein: instead the inner
details are only progressively revealed. However, as shown by Czerminski and Elber,\cite{Czerminski:90}
activated transitions between similar conformational structures confined to a
small set of minima (separated by barriers on the order of $k_B T$)
can occur at {\em room temperature} on a time scale of several
picoseconds. A much larger time scale is required for such transitions
at temperatures on the order of 1K. The insensitivity of $\gamma$ to
temperature therefore appears to rule out the first hypothesis: at the
low temperature end one would expect the protein to be confined to a
single minimum within the time scale of our simulations, and the
fractal dimension to be smaller as a consequence. Thus while
increased access to local minima is consistent with the growth of
$\gamma$ with time, it is inconsistent with the lack of temperature
dependence. We conclude that the fractal dimension measured in our
simulations is not a good probe of the {\em global} potential energy
landscape topology.

\begin{figure}
\hspace{6em}
\vspace{2em}
\epsfysize=6cm
\epsfxsize=11cm
\epsffile{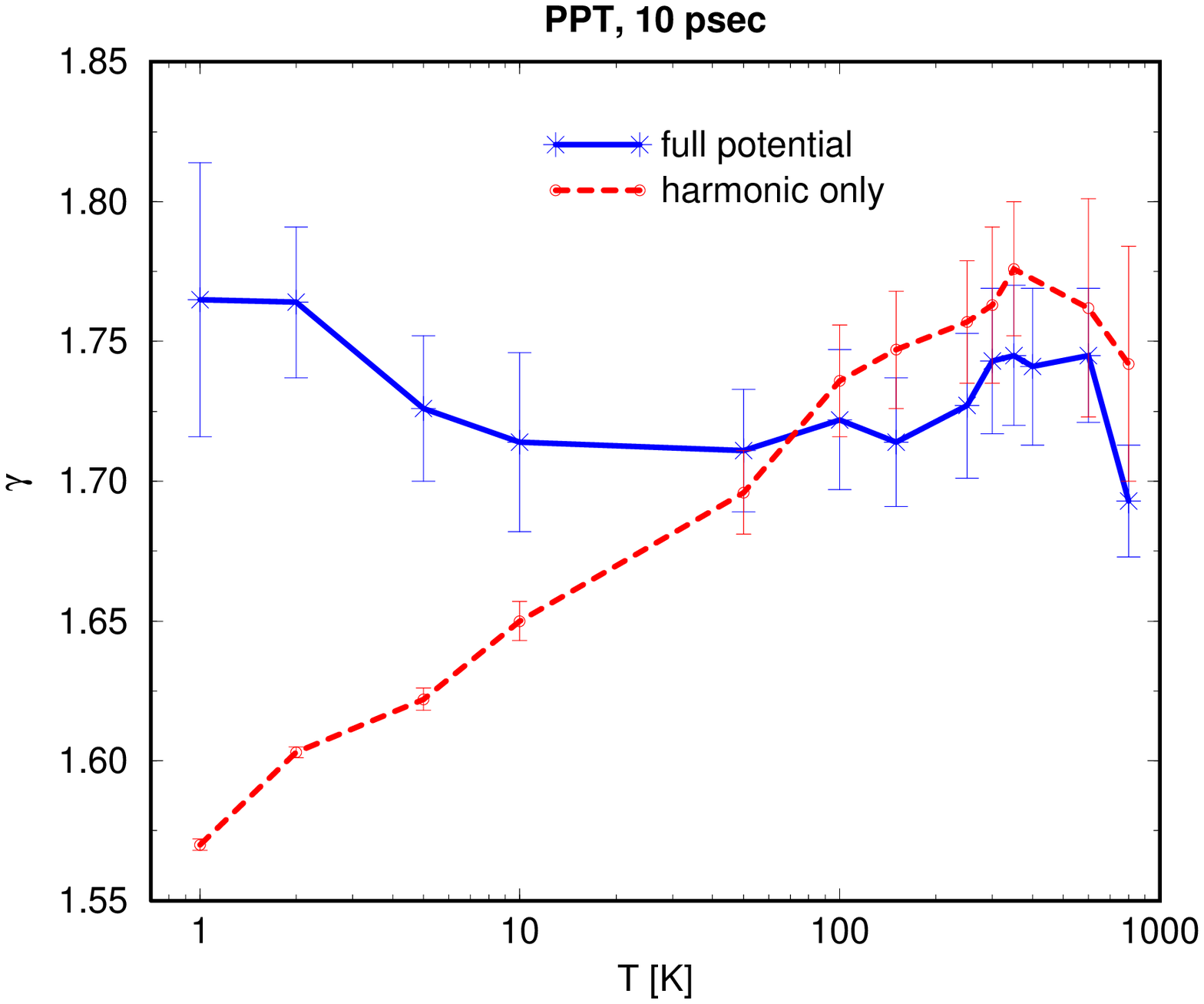}
\caption{10psec simulation results of PPT with the full
potential (solid line) and just the harmonic part (dashed line).}
\label{fig:12}
\end{figure}

This brings us to the second hypothesis which, as will be argued
shortly, is consistent 
with {\em both} the observed time and temperature
dependencies. According to this hypothesis, the primary factor
determining the fractal dimension of the potential energy curves is
the self-generated {\em dynamical noise}, namely the very broad frequency spectrum appearing in
the dynamics. This spectrum and the resulting self-generated noise is a consequence
of the {\em anharmonicity} in the potential, as is demonstrated
convincingly in Fig.\ref{fig:12}. Shown in this figure, along with the
full 10psec simulation of PPT at 300K, are the results of the same
simulations {\em with all anharmonic interactions switched off}: terms
$V_{\rm vdW}$, $V_{\rm electrostatic}$ and $V_{\rm torsion}$ are
missing from the total potential energy $V$ of Eq.(\ref{eq:V}). The
effect is dramatic: there is a strong decrease of $\gamma$ as the
temperature is lowered. This shows that anharmonicity is of paramount
importance in determining $\gamma$, at least up to 100K.

\begin{figure}
\hspace{6em}
\vspace{2em}
\epsfysize=6cm
\epsfxsize=11cm
\epsffile{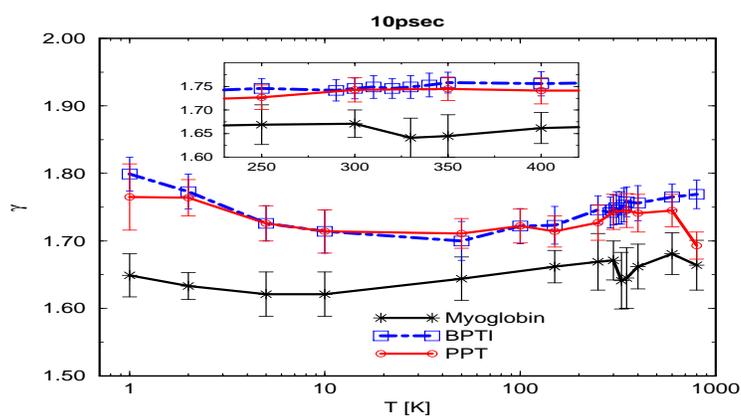}
\caption{Comparison among myoglobin, BPTI and PPT for 10psec
simulations.
}
\label{fig:7}
\end{figure}

\begin{figure}
\hspace{6em}
\vspace{2em}
\epsfysize=6cm
\epsfxsize=11cm
\epsffile{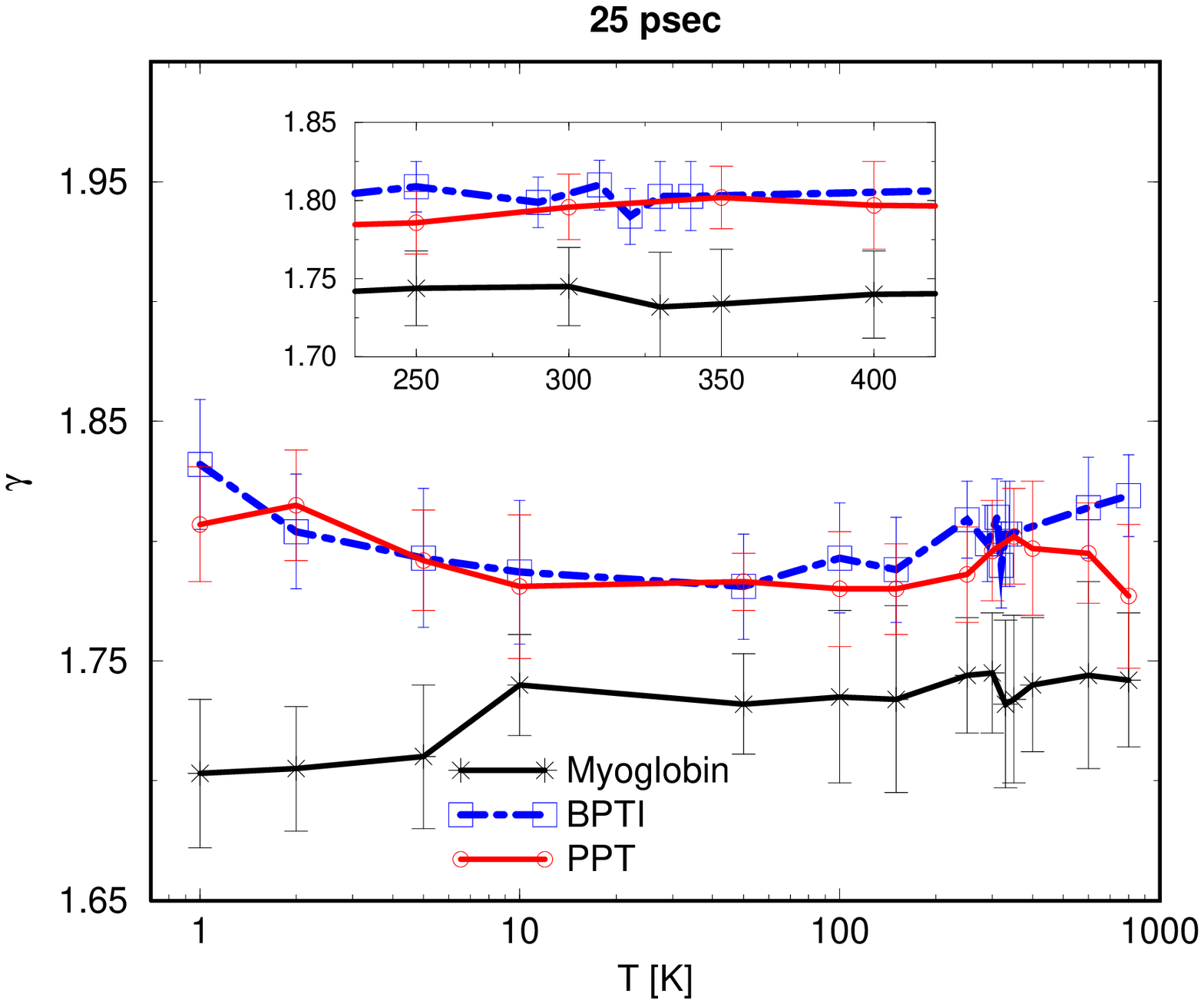}
\caption{Comparison among myoglobin, BPTI and PPT for 25psec
simulations. 
}
\label{fig:8}
\end{figure}

The mechanism
we propose is thus as follows: $V(t)$ is a sum over many anharmonic
terms, the number of which is proportional to the number of atoms in
the protein. These anharmonic terms create {\em overtones}, which
effectively fill the frequency spectrum and create a
quasi-continuum. Now, on the one hand, as time evolves (at constant
temperature) both very low 
and very high (previously inactive) frequency components 
become activated: the low frequencies are only sampled on long times, whereas the high frequencies correspond to highly energetic
modes which must await energy transfer before equilibration sets in. On the other hand, as temperature increases (at
constant $\tau$) intermode coupling causing energy transfer from rigid modes to soft modes is
facilitated, and more frequency components are activated. Thus in both
cases one expects an increase in dynamical noise, with a corresponding
increase in the fractal dimension of the signal. This accounts for the
behavior seen in Figs.\ref{fig:4}-\ref{fig:6}. Unfortunately we have
not been able to study whether the fractal dimension converges on a
limiting value upon full equilibration (at least as a function of temperature this does not
seem to be the case -- see Figs.\ref{fig:7} and \ref{fig:8}), since
this requires prohibitively long simulations.\cite{comment} 

The behavior of the harmonic case observed in
Fig.\ref{fig:12} might seem curious in light of these arguments: after
all in the purely harmonic case there is no intermode coupling, nor
any dynamical noise. However, we recall that due to the use of
non-Cartesian coordinates [see Eqs.(\ref{eq:bond}), (\ref{eq:angle})
and (\ref{eq:improper})], a certain {\em residual degree of anharmonicity}
remains with respect to the Cartesian components. At very low
temperatures vibrations and rotations occur very close to their
equilibrium values and the anharmonic effects are barely noticeable. Only
as the temperature rises do these vibrations and rotations start to deviate
significantly from their equilibrium values and gain
anharmonicity. This explains the rise of $\gamma$ in the ``harmonic'' case. In
contrast, the other terms [Eqs.(\ref{eq:vdW}), (\ref{eq:es})
and (\ref{eq:torsion})] make a strong anharmonic contribution at
all temperatures, which accounts for the relative constancy of $\gamma$
in this case.

The different behavior of PPT and BPTI {\it vs} myoglobin can be
attributed to the {\em size} of these proteins. Myoglobin
has 156 residues, whereas PPT and BPTI only have 41
and 60 residues respectively. Therefore at any given temperature it
takes longer to activate 
the overtones that give rise to the self-generated noise in myoglobin than in either PPT or BPTI, which
explains the lower fractal dimension of myoglobin in
Figs.\ref{fig:7},\ref{fig:8}. The proximity in size between PPT and
BPTI, on the other hand, is the reason for their quantitatively
similar behavior. A related argument makes clear why
deletion of the heavy heme group from myoglobin yields a higher
$\gamma$ (Fig.\ref{fig:4}): without heme the energy redistribution among
the modes occurs faster.

Also the curious similarity between NaCl and PPT can be attributed to
anharmonic effects: NaCl, with its purely Coulombic interactions, is of
course strongly anharmonic, even more so than PPT. On the other hand
poly-alanine is more harmonic, so as expected it has lower $\gamma$ values.

It is thus seen that the ``anharmonicity'' hypothesis is capable of
accounting for all of the salient features present in our results for
the fractal dimension of the $V(t)$ curves. The role of multiple
minima and ruggedness of the energy landscape seems to be more
restricted: it is apparently not the global topology which is measured
by $\gamma$. Nevertheless, it should be noted that {\em different}
multiple minima will generally lead to a denser filling of the
frequency spectrum or to the appearance of bands, which may overlap
and thus induce additional intermode coupling.

An important remaining issue concerns the {\em value} of $\gamma$
observed in our simulations: definitely greater than 1.5 (except
perhaps for the $\tau \!=\!$ 100fsec simulation,
Fig.\ref{fig:11}). {\it A priori} it is not clear what the lower limit of $\gamma$ for polymer-like systems
subject to the usual dynamics (like Newtonian or Langevin) should
be. A plausible lower limit of 1.5 is derived in
the appendix for the Rouse model. The Rouse model considers a purely harmonic
set of beads under the influence of a Gaussian random force. In our
simulations (in the ``no-equilibration'' -- microcanonical mode) there
is no randomness except for numerical truncation 
errors and ``randomness'' due to chaos caused by the anharmonic
terms. However, we find $\gamma \!\approx\! 1.5$ as the result of our ``harmonic'' simulations at
very low temperature (Fig.\ref{fig:12}). As argued above, this actually
corresponds to the low-anharmonicity limit, so is in agreement with
the result for the Rouse model, provided the truncation errors and/or the chaotic
``randomness'' can be modelled as a
Gaussian random force acting on the protein (this is a speculation). Increasing the degree of anharmonicity can
only increase the fractal dimension of the signal, by the mechanisms discussed
before. These arguments then show that the Rouse model
result could thus serve as a {\em lower bound} to the $\gamma$ values
found in our simulations.

\section{Summary and Conclusions}
\label{conclusions}

In this work we have studied the fractal properties of the total
potential energy $V$ as a function of time $t$ for several realistic
proteins. We performed simulations at a wide range of temperatures and
a number of time scales. Our main conclusion is that the fractal
behavior seems to be caused by the self-generated dynamical noise due to intermode coupling and
equilibration promoted by anharmonic effects. There are 
universal aspects to this behavior, as exemplified by the similarity
between totally different systems such as PPT and NaCl. Thus it cannot
serve to characterize individual proteins and their energy landscape,
but is rather a property common to all sufficiently large and
anharmonic systems. We believe, however, that probing the fractal
aspects of protein potential energy landscapes is a useful approach to
the characterization of the ruggedness of these landscapes. To achieve
this, it would be very interesting to examine $V$ as a function of
appropriate {\em spatial} coordinates, rather than $t$. The question
remains whether there are other ways leading to useful geometrical
characterizations of the energy landscape, e.g., identification of
multiple minima and maxima by real and imaginary
frequencies.\cite{Straub:94}

{\it Note added}: While this paper was in the final stages of
preparation we became aware of an article by Garc\'{i}a et al.\cite{Garcia:97}
The authors concluded that the time dependence of the mean-square
displacement of crambin around the crystal structure is described by
an effective fractal exponent $\gamma \approx 1.60$ (for $t$
long). This result was interpreted as indicative of multi-basin
dynamics which would be typical in a rugged energy landscape .\cite{Straub:94}

\section*{Acknowledgements}
We would like to thank Dr. Susan K. Gregurick and Dr. Adrian Roitberg
for invaluable technical assistance. DAL would
like to thank the Institute for Physical
Science and Technology at the University of Maryland, College Park,
for its hospitality during a visit in which part of this work was
carried out. We are grateful to Dr. Angel Garc\'{i}a for bringing
Ref.[\onlinecite{Garcia:97}] to our attention. This work was supported
in part by a   
grant from the National Science Foundation (NSF-CHE96-29845) to DT.

\appendix
\label{Rouse}
\section*{}

\begin{center}
{\bf AN ANALYTICALLY SOLVABLE EXAMPLE: THE ROUSE MODEL}\\
\end{center}

It is useful to have a simple reference model in the background
to compare the numerical results of Sec.\ref{results} with. To this
end, we consider the Rouse Model (RM), which is standard in polymer
theory. It describes a 
polymer as a set of $N$ beads located at $({\bf R}_1, {\bf R}_2, ...,
{\bf R}_N)$ connected along a chain, where each bead ($n$)
undergoes Brownian motion under the influence of random force ${\bf
f}_n(t)$. It should be remarked that this is essentially different
from our simulations, in which no randomness (except for numerical
truncation errors and chaotic ``randomness'' due to anharmonicity) is present. Nevertheless, it useful to study the RM in
order to see analytically how a fractal $V(t)$ can come about.

The excluded volume and hydrodynamic
interactions are disregarded in the RM. The dynamics is generally
described by the Langevin equation:\cite{Doi:86} 

\begin{equation}
{\partial \over {\partial t}} {\bf R}_n(t) = \sum_m {\bf H}_{nm} \cdot
\left( {\bf f}_m(t) - {{\partial U} \over {\partial {\bf R}_m}}
\right) + {1 \over 2} k_B T \sum_m {\partial \over {\partial {\bf
R}_m}} \cdot {\bf H}_{nm} .
\label{eq:general-L}
\end{equation}

\noindent In the RM the mobility tensor is  assumed to be
isotropic:

\begin{equation}
{\bf H}_{nm} = {{\bf I} \over \zeta} \delta_{nm} ,
\label{eq:H}
\end{equation}

\noindent and the interaction potential is taken to be harmonic:

\begin{equation}
U = {k \over 2} \sum_{n=2}^N ({\bf R}_n - {\bf R}_{n-1})^2 ,
\label{eq:U}
\end{equation}

\noindent so that Eq.(\ref{eq:general-L}) reduces to

\begin{eqnarray}
\zeta {{d {\bf R}_n} \over dt} & = & -k (2{\bf R}_n - {\bf R}_{n+1} - {\bf
R}_{n-1}) + {\bf f}_n \:\:\: (n=2,3,...,N-1) \\
\zeta {{d {\bf R}_1} \over dt} & = & -k ({\bf R}_1 -{\bf R}_2) + {\bf f}_1
\\
\zeta {{d {\bf R}_N} \over dt} & = & -k ({\bf R}_N -{\bf R}_{N-1}) + {\bf
f}_N .
\end{eqnarray}

\noindent The force and the potential define a time scale through

\begin{equation}
\tau \equiv {\zeta \over k} .
\label{eq:tau}
\end{equation}

\noindent It is convenient to consider the last set of equations in the
limit where $n$ is taken as a continuous variable:

\begin{eqnarray}
\zeta {{\partial {\bf R}_n} \over {\partial t}} &=& k {{\partial^2 {\bf
R}_n} \over {\partial n^2}} + {\bf f}_N \\  \nonumber
{{\partial {\bf R}_{n=0}} \over {\partial n}} &=& {{\partial {\bf R}_{n=N}}
\over {\partial n}} = 0 .
\label{eq:cont}
\end{eqnarray}

\noindent The random forces are assumed to be Gaussian, i.e.,
delta-function uncorrelated:

\begin{eqnarray}
\langle {\bf f}_n(t) \rangle &=& 0 \\ \nonumber
\langle {\bf f}_n(t){\bf f}_m(t') \rangle &=& 2\zeta k_B T \delta(n-m)
\delta(t-t') .
\label{eq:f}
\end{eqnarray}

\noindent In the continuum limit the potential energy becomes:

\begin{equation}
U = {k \over 2} \int_0^N dn\: \left( {{\partial {\bf R}_n} \over
{\partial n}} \right)^2 .
\label{eq:cont-U}
\end{equation}

\noindent The last three sets of equations define the continuum RM.

\subsection{Normal Coordinates}

We are interested in calculating the fractal scaling of the potential
energy $U(t)$ defined in Eq.(\ref{eq:cont-U}). For this purpose it is
convenient to transform to normal coordinates. It can be shown that in
terms of the coordinates

\begin{equation}
{\bf X}_p \equiv {1 \over N} \int_0^N dn\: \cos(p \pi n/N) {\bf R}_n(t)
\:\:\:\:\: p=0,1,2,...
\label{eq:normal-coords}
\end{equation}

\noindent the Langevin equation (\ref{eq:cont}) becomes

\begin{equation}
\zeta_p {d{\bf X}_p \over dt} = -k_p{\bf X}_p + {\bf f}_p
\label{eq:L-nc}
\end{equation}

\noindent where

\begin{eqnarray}
\zeta_0 &=& N\zeta \\
\zeta_p &=& 2N\zeta \:\:\:\: {\rm for} \:\:\:\: p=1,2,... \\
k_p &=& {{2\pi^2 k p^2} \over N} \:\:\:\: {\rm for} \:\:\:\: p=0,1,2,... 
\label{eq:defs}
\end{eqnarray}

\noindent and the forces satisfy

\begin{equation}
\langle {\bf f}_p(t) \rangle = 0 \\
\label{eq:fp1}
\end{equation}
\begin{equation}
\langle {\bf f}_p(t){\bf f}_q(t') \rangle = 2\zeta_p k_B T \delta{pq}
\delta(t-t') .
\label{eq:fp2}
\end{equation}

\noindent In this representation the random forces as well as the
motions of the ${\bf X}_p$'s are independent, so that the motion of
the polymer is decomposed into independent modes. Indeed, it is easily
verified that the formal solution to Eq.(\ref{eq:L-nc}) is

\begin{equation}
{\bf X}_p(t) = {1 \over \zeta_p} \int_{-\infty}^t dt' \:
e^{-(t-t')/\tau_p} {\bf f}_p(t') ,
\label{eq:sol}
\end{equation}

\noindent where

\begin{equation}
\tau_p \equiv {\tau_1 \over p^2} \:\:\: , \:\:\:\:\: \tau_1 \equiv
{\zeta_1 \over k_1} = {N^2 \over \pi^2} \tau .
\label{eq:taup}
\end{equation}

\noindent Finally, the inverse transform of Eq.(\ref{eq:normal-coords}) is

\begin{equation}
{\bf R}_n = {\bf X}_0+ 2\sum_{p=1}^{\infty} {\bf X}_p \cos(p \pi n/N) .
\label{eq:inverse}
\end{equation}

\subsection{Scaling of the Potential Energy}

We will now
present a calculation of the fractal scaling of the
potential energy in the RM, as defined in Eq.(\ref{eq:C}):\cite{comment2}

\begin{equation}
C_U(t) = \langle [ U(t) - U(0) ]^2 \rangle .
\label{eq:cU}
\end{equation}

\noindent Inserting the expression for ${\bf R}_n$ into
Eq.(\ref{eq:cont-U}) and expanding yields:

\begin{equation}
U = 2k \left( {\pi \over N} \right)^2 \sum_{p,q=1}^N p\,q {\bf X}_p(t)
\cdot {\bf X}_q(t) \int_0^N dn\: \sin(p \pi n/N) \sin(q \pi n/N) .
\label{eq:cont-U-1}
\end{equation}

\noindent The integral evaluates to ${N \over 2} \delta_{pq}$, so
that:

\begin{equation}
U = k {\pi^2 \over N} \sum_{p=1}^N p^2 {\bf X}^2_p(t) .
\label{eq:cont-U-2}
\end{equation}

\noindent Thus the fluctuation can be written as:

\begin{eqnarray}
C_U(t) &=& \left( {k\pi^2 \over N} \right)^2 \Bigl \langle \left[ \sum_{p=1}^N p^2
\left( {\bf X}^2_p(t) - {\bf X}^2_p(0) \right) \right]^2 \Bigl \rangle \\ \nonumber
&=& \left( {k\pi^2 \over N} \right)^2 \sum_{p,q=1}^N p^2\,q^2 \left[
\langle {\bf X}^2_p(t) {\bf X}^2_q(t) \rangle -2 \langle {\bf
X}^2_p(t) {\bf X}^2_q(0) \rangle + \langle {\bf X}^2_p(0) {\bf
X}^2_q(0) \rangle \right] .
\label{eq:cU-1}
\end{eqnarray}

\noindent Inserting the formal solution for the normal coordinates,
Eq.(\ref{eq:sol}), the averages can be calculated from:

\begin{eqnarray}
\langle {\bf X}^2_p(t) {\bf X}^2_q(t') \rangle = 
{1 \over {\zeta_p^2
\zeta_q^2}} \int_{-\infty}^t dt_1 \int_{-\infty}^t dt_2
\int_{-\infty}^{t'} dt_3 \int_{-\infty}^{t'} dt_4 \: \times \nonumber \\
e^{-(2t-t_1-t_2)/\tau_p} e^{-(2t'-t_3-t_4)/\tau_q}
\langle {\bf f}_p(t_1) \cdot {\bf f}_p(t_2) \: {\bf f}_q(t_3) \cdot {\bf
f}_q(t_4) \rangle
\label{eq:trouble}
\end{eqnarray}

\noindent To proceed the averages over the forces must be
evaluated. This can be done with the help of Wick's theorem:\cite{Doi:86}

\begin{equation}
\langle x_{n_1} x_{n_2} \cdots x_{n_{2p}} \rangle = \sum_{\rm all\:
pairings} \langle x_{m_1} x_{m_2} \rangle \cdots \langle x_{m_{2p-1}}
x_{m_{2p}} \rangle
\label{eq:wick}
\end{equation}

\noindent where the $x_n$'s are Gaussian random variables. Employing
the orthogonality relations from Eq.(\ref{eq:fp2}) thus yields:

\begin{eqnarray}
\langle {\bf f}_p(t_1) \cdot {\bf f}_p(t_2) \: {\bf f}_q(t_3) \cdot {\bf
f}_q(t_4) \rangle = \nonumber \\
(2k_B T)^2 \left[
\zeta_p \zeta_q \delta(t_1-t_2)\delta(t_3-t_4) + 
\zeta_p^2 \delta_{pq}^2 \delta(t_1-t_3)\delta(t_2-t_4) + 
\zeta_p^2 \delta_{pq}^2 \delta(t_1-t_4)\delta(t_2-t_3) \right]
\label{eq:deltas}
\end{eqnarray}

\noindent In addition we need the rule

\begin{equation}
\int_{-\infty}^t dt_2 \: f(t-t_2) \delta(t_2-t_1) = f(t-t_1) \Theta(t-t_1)
\label{eq:rule}
\end{equation}

\noindent where $\Theta(x)$ is the Heavyside step-function. Combining
these results, the delta functions simplify the expression in
Eq.(\ref{eq:trouble}) as follows:

\begin{eqnarray}
\langle {\bf X}^2_p(t) {\bf X}^2_q(t') \rangle =
(2k_B T)^2 \left[ {1 \over {\zeta_p \zeta_q}}
\int_{-\infty}^t dt_1 \: e^{-(t-t_1)(1/\tau_p + 1/\tau_1)}
\int_{-\infty}^{t'} dt_3 \: e^{-(t'-t_3)(1/\tau_p + 1/\tau_1)} + \right. \nonumber \\
\left. 2 {\delta_{pq} \over {\zeta_p^2}} \left( \int_{-\infty}^{\min
(t,t')} dt_1 \: e^{-(t+t'-2t_1)/\tau_p} \right)^2 \right]
= \nonumber \\
(2k_B T)^2 \left[ \left({{\tau_p \tau_q} \over {\tau_p +
\tau_q}}\right)^2 {1 \over {\zeta_p \zeta_q}} + 2\delta_{pq}
{{\tau_p^2} \over {\zeta_p^2}} e^{2(2\min(t,t')-t-t')/\tau_p} \right] .
\label{eq:ave1}
\end{eqnarray}

\noindent Inserting this into Eq.(\ref{eq:cU-1}) yields:

\begin{equation}
C_U(t) = \left( {k\pi^2 \over N} \right)^2 (2k_B T)^2 \sum_{p,q=1}^N
p^2\,q^2 \left[ \left({{\tau_p \tau_q} \over {\tau_p +
\tau_q}}\right)^2 {1 \over {\zeta_p \zeta_q}} + 
 4 {{\tau_p^2} \over {\zeta_p^2}} \left( 1 - e^{2t/\tau_p} \right)
\delta_{pq} \right] .
\label{eq:cU-2}
\end{equation}

\noindent Using the definitions of the various parameters appearing
here the final result becomes:

\begin{equation}
C_U(t) =  (k_B T)^2 \left[ \sum_{p,q=1}^N
{1 \over {p^2 + q^2}} + 4 \sum_{p=1}^N \left( 1 - e^{-2\pi^2 {t \over
\tau} {p^2 \over N^2}} \right) \right] .
\label{eq:cU-3}
\end{equation}

\noindent The time dependence of $C_U(t)$ for large but finite $N$
arises from the term $\sum_{p=1}^N \exp(-2\pi^2 t p^2/\tau N^2 )$. To derive an expression for $\gamma$, it is useful to
approximate the above sums by integrals. Defining $x=p/N$, $y=q/N$, we
have:

\begin{equation}
C_U(t) \approx  (k_B T)^2 \left[ 4N + {1 \over N^2} \int_{1/N}^1 dx\:
\int_{1/N}^1 dy\: {1 \over {x^2 + y^2}} - 4 \int_{1/N}^1 dx\:
e^{-2\pi^2 {t \over \tau} x^2} \right] .
\label{eq:cU-4}
\end{equation}

\noindent The first integral is easily evaluated after a
transformation to polar coordinates, and the second corresponds to the
error-function (in the second integral the error in taking $1/N = 0$
is negligible):

\begin{equation}
C_U(t) \approx (k_B T)^2 \left[ 4N + 2\pi{\ln N \over N^2} -
2\sqrt{\pi} {{{\rm erf}\sqrt{\theta}} \over \sqrt{\theta}} \right]
\:\:\:\:\:, \:\:\: \theta \equiv 2\pi^2 t/\tau .
\label{eq:cU-5}
\end{equation}

\noindent In accordance with the preceding comments we next take the $t \ll \tau$ limit. Then,
using ${\rm erf}(x) \approx {2 \over \sqrt{\pi}} (x - {x^3 \over
{3\cdot 1! }} + {x^5 \over {5\cdot 2! }} - ...)$ to the first
non-trivial (third) order, we can write:

\begin{eqnarray}
C_U(t \ll \tau) \approx (k_B T)^2 \left[ 4N + 2\pi{\ln N \over N^2} -
4 (1-{1\over 3}\theta) \right] \nonumber \\
\sim (k_B T)^2 {8\pi^2 \over 3} {t \over \tau} \sim
t^{2(2-\gamma)} ,
\label{eq:cU-6}
\end{eqnarray}

\noindent where in the last line the time-independent terms were
ignored, and ultimately $\gamma = 3/2$. Thus the Rouse model has a
potential energy fluctuations which have a fractal dimension of 3/2,
just like a classical, fully correlated random walk.

\newpage

\vspace{1.em}

\vspace{1.em}

\vspace{1.em}

\vspace{1.em}

\vspace{1.em}

\vspace{1.em}

\vspace{1.em}

\vspace{1.em}

\vspace{1.em}

\vspace{1.em}

\vspace{1.em}

\vspace{1.em}


\small
\begin{thebibliography}{10}

\bibitem{Frauenfelder:94}
H. Frauenfelder and P. Wolynes, Physics Today {\bf 47},  58  (1994).

\bibitem{Wolynes:95}
{P.G. Wolynes, J.N. Onuchic and D. Thirumalai}, Science {\bf 267},  1619
  (1995).

\bibitem{Straub:94}
{J.E. Straub, A.B. Rashkin and D. Thirumalai}, J. Am. Chem. Soc. {\bf 116},
  2049  (1994).

\bibitem{Czerminski:90}
{R. Czerminski and R. Elber}, J. Chem. Phys. {\bf 92},  5580  (1990).

\bibitem{Elber:95}
{R. Elber, A. Roitberg, C. Simmerling, R. Goldstein, H. Li,
G. Verkhivker, C. Kaeser, J. Zhang and A. Utilsky}, Comp. Phys. Commun.
  {\bf 91},  159  (1995).

\bibitem{Berry}
{K.D. Ball, R.S. Berry, R.E. Kunz, Li Feng-Yin, A. Proykova and
D.J. Wales}, Science {\bf 271}, 963 (1996).

\bibitem{comment}
It is clear that the upper bound for $\gamma(\tau \rightarrow \infty)$
is 2, so  Eq.~(\ref{eq:gamma-tau}) can be valid only for short $\tau$.

\bibitem{Family:91}
{\em Dynamics of fractal surfaces}, edited by F. Family and T. Vicsek (World
  Scientific, Singapore, 1991).

\bibitem{Russ:94}
{J.C. Russ}, {\em {Fractal Surfaces}} ({Plenum Press}, {New York}, 1994).

\bibitem{Stanley}
{\em {On Growth and Form}}, No.~100 in {\em NATO ASI Ser. E}, edited by {H. E.
  Stanley, N. Ostrowsky} (Martinus Nijhoff, Dordrecht, 1986).

\bibitem{Thirumalai:96}
{D. Thirumalai and S.A. Woodson}, Acc. Chem. Res. {\bf 29}, 433 (1996).

\bibitem{Avnir:book}
{\em {The Fractal Approach to Heterogeneous Chemistry: Surfaces, Colloids,
  Polymers}}, edited by {D. Avnir} ({John Wiley \& Sons Ltd.}, Chichester,
  1992).

\bibitem{Mandelbrot}
{B. B. Mandelbrot}, {\em {The Fractal Geometry of Nature}} ({Freeman}, {San
  Francisco}, 1982).

\bibitem{Falconer}
{K. Falconer}, {\em {Fractal Geometry: Mathematical Foundations and
  Applications}} ({Wiley}, {Chichester}, 1990).

\bibitem{Pfeifer-Avnir:book}
{P. Pfeifer and M. Obert},  in {\em {The Fractal Approach to Heterogeneous
  Chemistry: Surfaces, Colloids, Polymers}}, edited by {D. Avnir} ({John Wiley
  \& Sons Ltd.}, Chichester, 1992).

\bibitem{Stanley:2}
{A.-L. Barab\'{a}si and H.E. Stanley}, {\em {Fractal Concepts in Surface
  Growth}} ({Cambridge University Press}, {Cambridge}, 1995).

\bibitem{me:PRE96}
{D.A. Hamburger-Lidar}, Phys. Rev. E {\bf 54},  354  (1996).

\bibitem{Dubuc}
{B. Dubuc, J.P. Quiniou, C. Roques-Carmes, C. Tricot and S.W. Zucker}, Phys.
  Rev. A {\bf 39},  1500  (1989).

\bibitem{Garcia:97}
{A.E. Garc\'{i}a, R. Blumenfeld, G. Hummer and J.A. Krumhansl},
Physica D {\bf 107}, 225 (1997).

\bibitem{Doi:86}
{Doi and Edwards}, {\em {The Theory of Polymer Dynamics}} ({Oxford University
  Press}, {New York}, 1986), p.91.

\bibitem{comment2}
It may be tempting to consider this fluctuation in the potential energy
as related to the specific heat $C_V$. However, that is actually quite a
different fluctuation: $k_B T^2 C_V = \langle (E - \langle E \rangle
)^2 \rangle$, where $E$ is the {\em total} energy, and the fluctuation
is time-independent, near equilibrium.


\end{thebibliography}
\end{document}